\documentclass[letter]{aa}

\usepackage{graphicx}
\usepackage{natbib}
\usepackage{hyperref}
\usepackage{txfonts}

\begin{document}

   \title{HS Hya about to turn off its eclipses}

   \titlerunning{HS Hya about to turn off its eclipses}

   \author{P. Zasche \inst{1}
            \and A. Paschke\inst{2}}

   \authorrunning{P. Zasche \& A. Paschke}

   \institute{Astronomical Institute, Faculty of Mathematics and Physics, Charles University Prague,
   CZ-180 00 Praha 8, V Hole\v{s}ovi\v{c}k\'ach 2, Czech Republic, \email{zasche@sirrah.troja.mff.cuni.cz}
   \and
   Weierstr 30, 8630 Rueti, Switzerland
   }

   \date{\today}

  \abstract
   {}
   {We aim to perform the first long-term analysis of the system HS Hya.}
   {We performed an analysis of the long-term evolution of the light curves of the detached eclipsing system
   HS Hya. Collecting all available photometric data since its discovery, the light curves were analyzed with a special
   focus on the evolution of system's inclination.}
   {We find that the system undergoes a rapid change of inclination. Since its discovery until today the system's
   inclination changed by more than 15$^\circ$. The shape of the light curve changes, and now the eclipses are
   almost undetectable. The third distant component of the system is causing the precession of the close orbit,
   and the nodal period is about 631~yr.}
   {New precise observations are desperately needed, preferably this year, because the amplitude of variations is
   decreasing rapidly every year. We know only 10 such systems on the whole sky at present.}


  \maketitle


\section{Introduction}

Eclipsing binaries are astronomical objects of high importance, especially owing to the possibility
of deriving the basic physical properties of these stars with high precision. This is mostly
because one can easily calculate the individual masses, semimajor axis, etc. if one knows the
inclination of the system and the radial velocity curves. However, in some systems the plane of the
orbit is moving slowly, and the radial velocity data have to be obtained at the same time as the
data for the light curve solution. Otherwise, the method yields incorrect results.

At present, we know only a few systems where the orbital plane is moving and the eclipsing light
curve had different shapes in different epochs. Six such systems were summarized by
\cite{2005Ap&SS.296..113M}. Moreover, three more systems show changes of minima depths, therefore
they are also suspected to undergo a precession of the orbits, these are V685~Cen
\citep{2004IBVS.5563....1M}, AH~Cep \citep{1989A&A...221...49D}, and V699 Cyg
\citep{1991IBVS.3667....1A}.

\section{The system HS Hya}

The eclipsing binary system HS Hya was discovered to be variable by \cite{1965IBVS..107....1S}, who
also classified the system as an Algol-type, but the orbital period given is incorrect.
\cite{1971ApJ...166..361P} measured the radial velocities (hereafter RV) of the system, and
analyzed the RV curves. The spectral type was derived to be F3-4, the correct orbital period is
given to be about 1.568024~days, and the mass ratio is about 0.96. However, the RV data were
obtained over a period of five years, from 1966 to 1970. Later, the complete light curve (hereafter
LC) was obtained in the Str\"omgren $uvby$ system by \cite{1975A&A....42..303G}. These data were
measured in 1972. The authors used the RV results from \cite{1971ApJ...166..361P} and their
inclination of about 85.3$^\circ$, which yielded a reliable picture of the system.

However, \cite{1997AJ....114.2764T} published a new finding about HS Hya. The analysis was based on
older data from photometry, RVs by \citeauthor{1971ApJ...166..361P}, but also Torres and coworker's
own new RV data, revealing that one more distant component is orbiting around the eclipsing pair.
The period of this body is about 190~days, and it is probably of the spectral type M0. Its light
contribution is quite low (below 1\%), but the RV residuals obtained by a cross-correlation clearly
show periodic modulation.

\begin{figure}
  \centering
  \includegraphics[width=88mm]{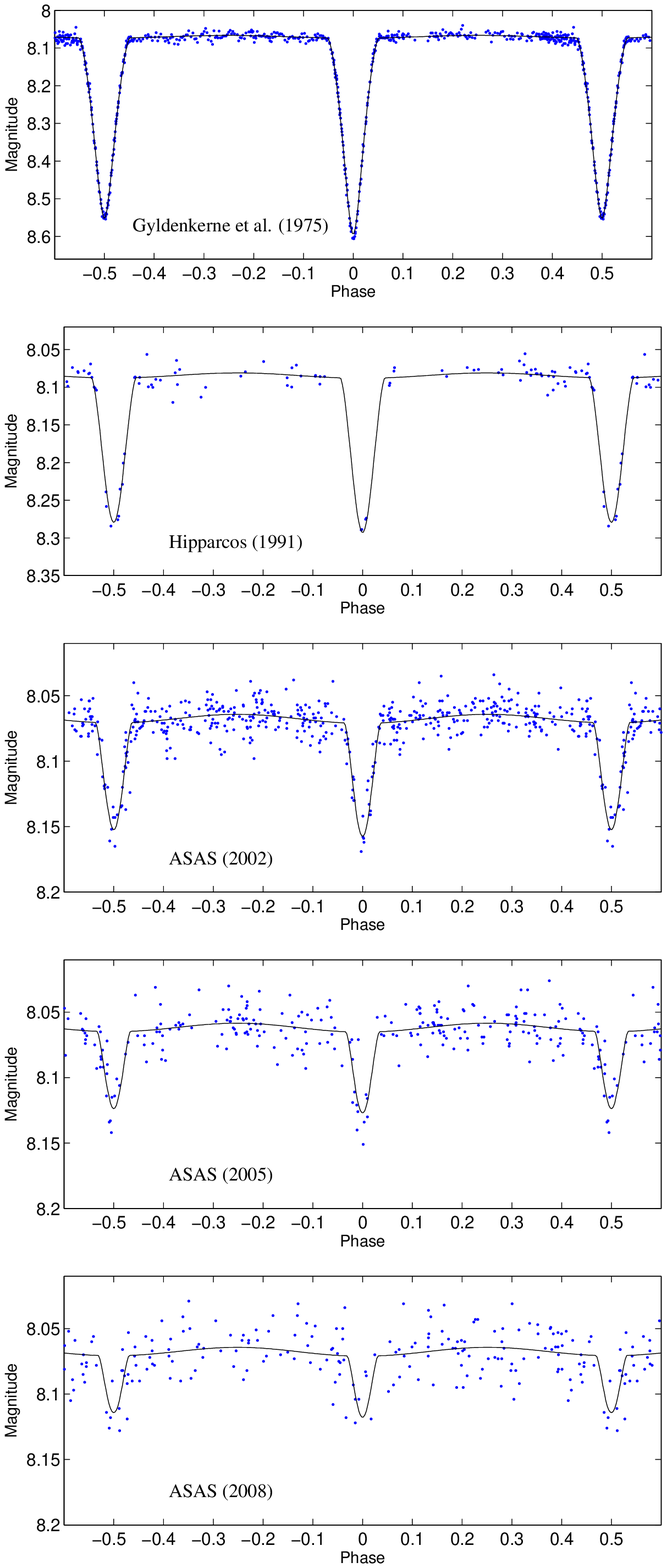}
  \caption{Available light curves of HS Hya in the $V$ filter. The changing depth of both minima is
  clearly visible. The three bottom figures were plotted with the same range in the y-axis.}
  \label{FigLCs}
\end{figure}

\section{The change of inclination}

The star has also been observed by the Hipparcos satellite \citep{HIP}. During its three-year
mission both minima were observed, but the coverage is only poor. However, after transformation
from $H_p$ to $V$ magnitude \citep{2001A&A...369.1140H}, it is clear that the depths of both minima
are much lower than in the LCs obtained 20~yr ago. We solved the Hipparcos LC with the same
parameters as given in \cite{1997AJ....114.2764T}. The {\sc PHOEBE} program \citep{Prsa2005} was
used, which is based on the Wilson-Devinney code, \cite{Wilson1971}. All relevant parameters were
fixed except for the inclination, see Table \ref{TabINCL}. The value of the third light was fixed
at a value of 0.4\% only, in agreement with the finding published by \cite{1997AJ....114.2764T}.

\begin{table}
 \centering
  \caption{The inclination as obtained from various light curves.}  \label{TabINCL}
  \begin{tabular}{c c c}
\hline
 Year & Inclination [deg]  & Reference \\
 \hline
  1964 & 88.9  $\pm$  1.1 & {\cite{1965IBVS..107....1S}} \\
  1972 & 85.30 $\pm$ 0.41 & {\cite{1975A&A....42..303G}} \\
  1991 & 79.83 $\pm$ 0.21 & {\cite{HIP}} \\
  2002 & 76.13 $\pm$ 0.15 & ASAS \\
  2005 & 75.19 $\pm$ 0.28 & ASAS \\
  2008 & 74.60 $\pm$ 0.50 & ASAS \\
 \hline
\end{tabular}
\end{table}

The star was also included into the photometric survey ASAS \citep{2002AcA....52..397P}. We divided
the whole data set into three parts and separately solved the light curves in 2002, 2005, and 2008.
The procedure of LC fitting was the same as for the Hipparcos data, and the results are given in
Table \ref{TabINCL}. As one can see from Fig.\ref{FigLCs}, the depths of the minima are still
decreasing.

Unfortunately, it is not easy to find other reliable photometry to do a similar analysis in
different time epochs. One of the limiting problems is the role of the filter, because the entire
abovementioned photometry can easily be transformed into the standard $V$ magnitudes. The
photometry from the automatic survey called "Pi of the sky" \citep{2005NewA...10..409B} is another
possibility, but this photometry is unfiltered, and therefore it is problematic to solve its light
curve. Moreover, it has fairly high scatter and covers a similar time span as the ASAS data.

We also tried to use the data from the discovery paper \citep{1965IBVS..107....1S}, but these are
only the photographic data and were not obtained in any standard photometric filter. Another
problem is that the original data are not available, only the phase plot, but this was constructed
with incorrect ephemerides. As one can see from the LCs published in \cite{1975A&A....42..303G},
the individual depths in different filters are quite similar to each other. Moreover, fixing the
other relevant parameters during the fitting process, we were able to construct a plot of minima
depth versus inclination. Using the eight dimmed data points from \cite{1965IBVS..107....1S}, we
were able to roughly derive the inclination from these data obtained in 1964, see Table
\ref{TabINCL}.

\begin{figure}
  \centering
  \includegraphics[width=89mm]{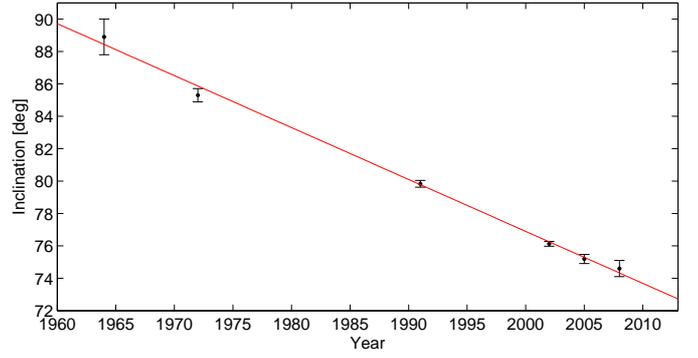}
  \caption{Inclination of the eclipsing binary with respect to the time.}
  \label{FigInclYr}
\end{figure}

The data from Table \ref{TabINCL} were used to construct the plot given in Fig.\ref{FigInclYr}.
Fitting these data points with a linear curve, one sees that the change of inclination is about
0.3$^\circ$ during one year. Therefore, the amplitude of photometric variations is decreasing
rapidly every year.

\section{The nodal period}

The precession effect of the close pair's orbit due to the distant third body was described
elsewhere, e.g. \cite{1975A&A....42..229S}. The nodal period can be computed from the equation
$$P_{\mathrm{nodal}} = \frac{4}{3} \left( 1+ \frac{M_1 + M_2}{M_3} \right) \frac{P_3^2}{P} (1-e_3^2)^{3/2}
\left( \frac{C}{G_2} \cos j \right)^{-1},$$ where subscripts 1 and 2 stand for the eclipsing binary
components, while 3 stands for the third distant body. The term ${G_2}$ stands for the angular
momentum of the wide orbit, and the $C$ is the total angular momentum of the system. However, the
problem of the unknown inclination of the wide orbit (which is included in the last term in
brackets) led us to use a different approach. \cite{1994A&A...284..853D} analyzed the system
IU~Aur, where a similar problem arose, hence one can also fit the term $\cos i$ with a sinusoidal
fit, following the equation
$$\cos i = \cos I \cdot \cos i_1 - \sin I \cdot \sin i_1 \cdot \cos (2\pi(t-t_0)/P_{\mathrm{nodal}}), $$ where $I$
is the inclination of the invariant plane against the observer's celestial plane, $i$ is the
inclination of the eclipsing binary, while $i_1$ is the inclination between the invariant plane and
the orbital plane of the eclipsing binary.

\begin{figure}
  \centering
  \includegraphics[width=89mm]{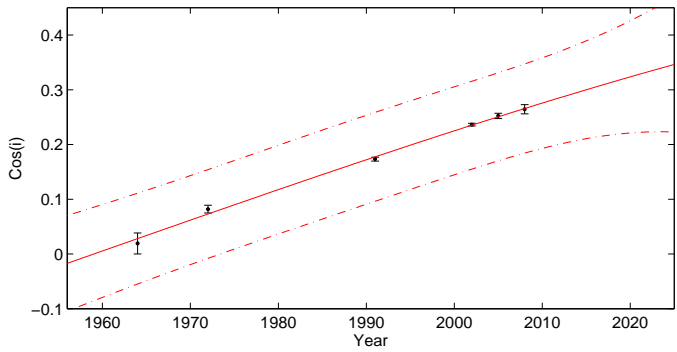}
  \caption{Fitting the $\cos(i)$ with the sinus term on available data, the confidence level of 95\% is
  shown with the dash-dotted line.}
  \label{FigCosIfit}
\end{figure}

A similar analysis was performed (see Fig.\ref{FigCosIfit}), but regrettably only a small part of
the nodal period is covered with data points nowadays. Fitting a sinusoidal curve, the nodal period
is about 631~yr, of which only about 1/14 is covered. The other two adjustable quantities were only
poorly constrained. New precise observations are urgently needed.

\section{Observational consequences}

HS~Hya belongs to a group of unique systems, and therefore more observations are needed to confirm
our hypothesis with higher confidence and to study its other physical properties. The amplitude of
eclipses is lower than 0.015~mag in $V$ filter at present, and is still decreasing. However, the
ellipsoidal variations outside the eclipse will remain even when the photometric eclipses
disappear. According to our model the photometric eclipses will stop in about 2022. Nevertheless,
detecting these shallow eclipses is problematic, especially from the northern hemisphere, owing to
the low declination of the star.

\begin{table*}
 \centering
  \caption{Known eclipsing binaries with changing minima depths.}  \label{TabSystems}
  \begin{tabular}{c c c c c c l}
\hline
 System  & Mag  &  Sp.   &  Eclipsing & Long   &  Nodal & Reference \\
         & $V$  & type   &  period    & period & period &        \\
         & [mag]&        &   [day]    & [day]  &   [yr] &  \\
 \hline
  IU Aur & 8.39 & B3Vnne &    1.8115  &  293.3&    330  & {\cite{2003A&A...403..675O}} \\
 V685 Cen& 8.85 & A0     &    1.1910  &  ?    &     ?   & {\cite{2004IBVS.5563....1M}} \\
  AH Cep & 6.88 & B0.5V  &    1.7747  &  ?    &     ?   & {\cite{1989A&A...221...49D}} \\
 V699 Cyg& 11.6 & B2     &    1.5515  &  ?    &     ?   & {\cite{1991IBVS.3667....1A}} \\
  SV Gem & 10.57& B3     &    4.0061  &  ?    &     ?   & {\cite{2001IBVS.5090....1G}} \\
  HS Hya & 8.07 & F5V    &    1.5680  &  190  &    631  & This study \\
  SS Lac & 10.12& B9V    &   14.4162  &  679  &    600  & {\cite{2001AJ....121.2227T}} \\
  AY Mus & 10.35& B9     &    3.2055  &  ?    &     ?   & {\cite{1975A&A....42..229S}} \\
  RW Per & 9.68 & A5Ve   &   13.1989  &  ?    &     ?   & {\cite{1992AJ....103..256O}} \\
 V907 Sco& 8.61 & B9V    &    3.7763  &  99.3 &    68   & {\cite{1999AJ....117..541L}} \\
 \hline
\end{tabular}
\end{table*}

Concerning the third body period, one can ask why the 190-day orbit was not discovered earlier via
analyzing the minima times via period variations. There exists a huge database of minima
observations (more than 100), but no variation was detected. This is because of the short period of
the third body and its low mass. This amplitude (see \citealt{Irwin1959}) resulted in about
1 minute only, 
which is comparable with the precision of individual times of minima observations.

\section{Discussion and conclusions}

Assembling all available systems with changing minima depths (see Table \ref{TabSystems}), one can
see how unique these systems are. We know only 10 such systems today and the nodal period was
derived in only four of them. Moreover, it seems that this effect was preferably observed in early
type systems (B and A spectral types), and HS~Hya is the first exception.

Our hypothesis of changing inclination would also slightly shift the physical parameters of
components as presented in \cite{1997AJ....114.2764T}. These authors assumed constant inclination
and used light curve and radial velocity data from different epochs. However, the time gap of more
than 20~years between photometry and spectroscopy yields a difference in inclination of about
7.8$^\circ$. This difference is able to shift the true stellar masses as computed from the term
$m\! \cdot \! \sin^3(i).$ The inclination in 1968 (when the RV data were obtained by
\citealt{1971ApJ...166..361P}) was about 87.15$^\circ$, while in 1992 (roughly the middle of the
time interval of radial velocities used in \citealt{1997AJ....114.2764T}) was about 79.3$^\circ$.
Using these values, the masses resulted in the values presented in Table \ref{Masses}.

\begin{table}
 \centering
  \caption{The masses as derived from different data sets/methods.}  \label{Masses}
\begin{tabular}{c c c c}
 \hline
     &  RV 1968  & RV 1992  & Torres et al.(1997) \\ \hline
 $M_1/M_\odot =$ & 1.319 & 1.307 & 1.255 \\
 $M_2/M_\odot =$ & 1.291 & 1.267 & 1.219 \\ \hline
\end{tabular}
\end{table}

These values agree much better with the predicted stellar masses of F4+F5 spectral types (i.e. with
the temperatures) for normal metallicity. Hence, the errors of masses as presented by
\cite{1997AJ....114.2764T} of about $0.007~M_\odot$ are too optimistic, because they neglect the
systematic effect described above. \cite{1997AJ....114.2764T} presented a perfect fit of the
derived values of $\log M$, $\log T$, and $\log g$ on the model isochrones. Likely an adjustment of
metallicity or age will be required to accommodate the new mass determinations.

The plot presented in Fig. \ref{FigCosIfit} needs to be spread in the next years. However, deriving
the inclination of HS~Hya when it stops having eclipses will be hard, because the ellipsoidal
variations have only low amplitude. On the other hand, the interferometry of the close pair would
solve this problem. Detecting the two eclipsing components via interferometry is difficult, but
worth trying. The system itself is relatively bright and the two eclipsing components have a
similar luminosity (i.e. magnitude difference close to zero). The computed angular separation of
the two eclipsing components is about 0.4~mas.

\begin{acknowledgements}
Pavel Mayer is acknowledged for a useful discussion and valuable advice. This work was supported by
the Czech Science Foundation grant no. P209/10/0715, by the grant UNCE 12 of the Charles University
in Prague, and also by the Research Programme MSM0021620860 of the Czech Ministry of Education.
This research has made use of the SIMBAD database, operated at CDS, Strasbourg, France, and of
NASA's Astrophysics Data System Bibliographic Services.
\end{acknowledgements}


\begin{thebibliography}{}

\bibitem[Azimov \& Zakirov(1991)]{1991IBVS.3667....1A} Azimov, A.~A., \& Zakirov, M.~M.\ 1991, IBVS, 3667, 1

\bibitem[Burd et al.(2005)]{2005NewA...10..409B} Burd, A., Cwiok, M., Czyrkowski, H., et al.\ 2005, New Ast, 10, 409

\bibitem[Drechsel et al.(1994)]{1994A&A...284..853D} Drechsel, H., Haas, S., Lorenz, R., \& Mayer,
P.\ 1994, \aap, 284, 853

\bibitem[Drechsel et al.(1989)]{1989A&A...221...49D} Drechsel, H., Lorenz, R., \& Mayer, P.\ 1989, \aap, 221, 49

\bibitem[Guilbault et al.(2001)]{2001IBVS.5090....1G} Guilbault, P.~R., Lloyd, C., \& Paschke, A.\
2001, IBVS, 5090, 1

\bibitem[Gyldenkerne et al.(1975)]{1975A&A....42..303G} Gyldenkerne, K., J{\o}rgensen, H.~E.,
\& Carstensen, E.\ 1975, \aap, 42, 303

\bibitem[Harmanec \& Bo{\v z}i{\'c}(2001)]{2001A&A...369.1140H} Harmanec, P., \& Bo{\v z}i{\'c},
H.\ 2001, \aap, 369, 1140

\bibitem[Irwin(1959)]{Irwin1959} {Irwin}, J.~B. 1959, \aj, 64, 149

\bibitem[Lacy et al.(1999)]{1999AJ....117..541L} Lacy, C.~H.~S., Helt, B.~E., \& Vaz, L.~P.~R.\ 1999, \aj, 117, 541

\bibitem[Mayer et al.(2004)]{2004IBVS.5563....1M} Mayer, P., Pribulla, T., \& Chochol, D.\ 2004, IBVS, 5563, 1

\bibitem[Mayer(2005)]{2005Ap&SS.296..113M} Mayer, P.\ 2005, \apss, 296, 113

\bibitem[Olson et al.(1992)]{1992AJ....103..256O} Olson, E.~C., Schaefer,
B.~E., Fried, R.~E., Lines, R., \& Lines, H.\ 1992, \aj, 103, 256

\bibitem[{\"O}zdemir et al.(2003)]{2003A&A...403..675O} {\"O}zdemir, S., Mayer, P., Drechsel, H.,
Demircan, O., \& Ak, H.\ 2003, \aap, 403, 675

\bibitem[Perryman et al.(1997)]{HIP} Perryman, M.~A.~C., Lindegren, L., Kovalevsky, J., et al.\ 1997, \aap, 323, L49

\bibitem[Pojmanski(2002)]{2002AcA....52..397P} Pojmanski, G.\ 2002, AcA, 52, 397

\bibitem[Popper(1971)]{1971ApJ...166..361P} Popper, D.~M.\ 1971, \apj, 166, 361

\bibitem[Pr{\v s}a \& Zwitter(2005)]{Prsa2005} Pr{\v s}a, A., \& Zwitter, T.\ 2005, ApJ, 628, 426

\bibitem[S\"oderhjelm(1975)]{1975A&A....42..229S} S\"oderhjelm, S. 1975, A\&A, 42, 229

\bibitem[Strohmeier et al.(1965)]{1965IBVS..107....1S} Strohmeier, W.,
Knigge, R., \& Ott, H.\ 1965, IBVS, 107, 1

\bibitem[Torres et al.(1997)]{1997AJ....114.2764T} Torres, G., Stefanik,
R.~P., Andersen, J., et al.\ 1997, \aj, 114, 2764

\bibitem[Torres(2001)]{2001AJ....121.2227T} Torres, G.\ 2001, \aj, 121, 2227

\bibitem[Wilson \& Devinney(1971)]{Wilson1971} Wilson, R.~E., \& Devinney, E.~J.\ 1971, ApJ, 166, 605

\end{thebibliography}
\end{document}